\documentclass{elsart}
\usepackage{epsfig}
\usepackage{amssymb}

\begin{document}

\begin{frontmatter}
\title{Underlying Dynamics of Typical Fluctuations of an Emerging Market Price Index: The Heston Model from Minutes to Months }

\author[1]{Renato Vicente}\ead{rvicente@usp.br} \author[2]{, Charles M. de Toledo} \author[3]{, Vitor B.P. Leite}
\author[4]{and Nestor Caticha}

\address[1]{Escola de Artes, Ci{\^e}ncias e Humanidades, Universidade de S\~ao Paulo, Parque Ecol\'ogico do Tiet\^e, 03828-020, S\~ao Paulo-SP, Brazil}
\address[2]{BOVESPA - S\~ao Paulo Stock Exchange, R. XV de Novembro, 275, 01013-001 S\~ao Paulo - SP, Brazil}
\address[3]{Dep. de F{\'\i}sica, IBILCE, Universidade Estadual Paulista, 15054-000 S\~ao Jos\'e do Rio Preto - SP, Brazil}
\address[4]{Dep. de F{\'\i}sica Geral, Instituto de F{\'\i}sica, Universidade de S\~ao Paulo, Caixa Postal 66318, 05315-970 S\~ao Paulo - SP, Brazil}

\begin{abstract}
We investigate the Heston model with stochastic volatility and exponential tails as a model for the typical price fluctuations of the Brazilian S\~ao Paulo Stock Exchange Index (IBOVESPA).  Raw prices are first corrected for inflation and a period spanning 15 years characterized by memoryless returns is chosen for the analysis. Model parameters are estimated  by observing volatility scaling and correlation properties. We show that the Heston model with at least two time scales for the volatility mean reverting dynamics satisfactorily describes price fluctuations ranging from time scales larger than 20 minutes to 160 days. At time scales  shorter than 20 minutes  we observe autocorrelated returns and power law tails incompatible with the Heston model. Despite major regulatory changes, hyperinflation and currency crises experienced by the Brazilian market in the period studied, the general success of the description provided  may be regarded as an evidence for a general underlying dynamics of price fluctuations at intermediate mesoeconomic time scales well approximated by the Heston model. We also notice that the connection between the Heston model and Ehrenfest urn models could be exploited for  bringing  new insights into  the microeconomic market mechanics.    
\end{abstract}


\end{frontmatter}

\section{Introduction}
\label{intro}
In the last decades the quantitative finance community has devoted much attention to the modeling of asset returns having as a major drive the improvement of pricing techniques \cite{fouque} by employing stochastic volatility models  ameanable to analytical treatment such as Hull-White \cite{hullwhite}, Stein-Stein \cite{steinstein} and Heston \cite{heston} models.  

Despite differences in methods and emphasis,  the cross fecundation between Economics and Physics, which dates back to the early nineteenth century (see \cite{roehner} and \cite{mirowski}), has intensified recently \cite{stanley}. Following the statistical physics approach, substantial effort has been made to find microeconomic models capable of reproducing a number of recurrent features of financial time series such as: returns aggregation (probability distributions at any time scale) \cite{yakovenko,silva}, volatility clustering \cite{engle}, leverage effect (correlation between returns and volatilities) \cite{bouchaudprl,perello}, conditional correlations \cite{boguna,lebaron} and fat tails at very short time scales \cite{stanley}. A central feature of economical phenomena is the prevalence of intertwined dynamics at several time scales. In very general terms, one could divide the market dynamics onto three broad classes: the microeconomic dynamics at the time scales of books of orders and price formation, the mesoeconomic dynamics at the scales of oscillations in formed prices due to local supply and demand and the macroeconomic dynamics at the scales of aggregated economy trends. 

The literature on empirical finance have emphasized the multifractal scaling \cite{dimateo} and the power law tails of price fluctuations. However it has been shown \cite{lebaron1} that very large data sets are required in order to distinguish between a multifractal and power law tailed process and a stochastic volatility model. In this paper we, therefore, deal with the mesoeconomic dynamics as it would be described by the Heston model with stochastic volatility and exponential tails.     

Recently, a semi-analytical solution for the Fokker-Planck equation describing the distribution of log-returns  in the Heston model has been proposed \cite{yakovenko}. The authors were able to show a satisfactory agreement between  return distributions of a number of developed market stock indices and the model for time scales spanning a wide interval ranging from 1 to 100 days. More recently, the same model has  also been employed to describe single stocks intraday fluctuations with relative success \cite{yakovenko2}.  

In this  paper we show evidence that the Heston model is also capable of describing the fluctuation dynamics of an emerging market price index, the Brazilian S\~ao Paulo Stock Exchange Index (IBOVESPA).
We employ for the analysis 37 years of data since IBOVESPA inception in January, 1968 and approximately four years of intraday data as well. In this period the Brazilian economy has experienced periods of  political and economical unrest, of hyperinflation, of currency crises and of  major regulatory changes. These distinctive characteristics make the IBOVESPA an interesting ground for general modelling and data analysis and for testing the limits of description of the Heston model.   

This paper is organized as follows. The next section surveys the Heston model, its semi-analytical solution and its connection to Ehrenfest urn models. Section \ref{sec:preprocessing} discusses the pre-processing necessary for isolating the fluctuations to be described by the Heston model from exogenous effects. Section \ref{sec:data_analysis} describes the data analysis at low (from daily fluctuations) and high (intraday fluctuations) frequencies. Conclusions and further directions are presented in Section \ref{sec:conclusions}.

\section{The Heston Model}
\label{heston}

\subsection{Semi-analytical solution}
The Heston Model describes the dynamics of stock prices $S_t$ as a geometric Brownian motion with volatility given by a Cox-Ingersoll-Ross (or Feller) mean-reverting dynamics. In the It\^o differential form the model reads: 
\begin{eqnarray}
\label{eq_heston}
dS_t &=& S_t\,\mu_t dt\;+\;S_t\sqrt{v_t}\;dW_0(t)\\
dv_t &=& -\gamma\left[v_t-\theta\right]dt\;+\;\kappa\sqrt{v_t}\;dW_1(t),\nonumber 
\end{eqnarray}
where $v_t$ represents the square of the volatility and $dW_j$ are Wiener processes with:
\begin{eqnarray}
\label{eq_Wiener}
\langle dW_j(t) \rangle &=& 0,\nonumber\\
\langle dW_j(t)\, dW_k(t^{\prime}) \rangle &=& 
\delta(t-t^{\prime})\left[\delta_{jk}\,dt+(1-\delta_{jk})\rho\,dt \right]
\end{eqnarray}

The volatility reverts towards a macroeconomic long term mean squared volatility $\theta$ with relaxation time given by $\gamma^{-1}$, $\mu_t$ represents a drift at macroeconomic scales, the coefficient $\sqrt{v_t}$ prevents negative volatilities and $\kappa$ regulates the amplitude of volatility fluctuations.

As we are mainly concerned with price fluctuations, we simplify equation (\ref{eq_heston}) by introducing log-returns in a window $t$ as $r(t)=\ln(S(t))-\ln(S(0))$. Using Ito's lemma and changing variables by making $ x(t)=r(t)-\int_0^tds\,\mu_s$ we  obtain a detrended version of the return dynamics that reads:
\begin{equation}
\label{eq_detrended}
dx=-\frac{v_t}{2}\,dt+\sqrt{v_t}\,dW_0.
\end{equation}

The solution of the Fokker-Planck equation (FP) describing the unconditional distribution of log-returns was obtained by Dr$\breve{a}$gulescu and Yakovenko \cite{yakovenko} yielding:
\begin{equation}
\label{px4}
P_t(x)=\int_{-\infty}^{+\infty}\frac{dp_x}{2\pi}\;e^{ip_xx+\alpha \phi_t(p_x)},
\end{equation}
where
\begin{eqnarray}
\label{eq_Omega}
\alpha&=&\frac{2\gamma\theta}{\kappa^2}\\
\phi_t(p_x)&=&\frac{\Gamma t}{2} - \ln\left[\cosh\left(\frac{\Omega t}{2}\right)+ \frac{\Omega^2-\Gamma^2+2\gamma\Gamma}{2\gamma\Omega}\sinh\left(\frac{\Omega t}{2}\right)\right]\\
\Omega&=&\sqrt{\Gamma^2+\kappa^2\left(p_x^2-ip_x\right)}\\
\Gamma&=&\gamma+i\rho\kappa p_x. 
\end{eqnarray}
This unconditional probability density has exponentially decaying tails and the following asymptotic form for  short time $t\ll \gamma^{-1}$:
\begin{equation}
\label{short_t}
P_t(x)=\frac{2^{1-\alpha}e^{-x/2}}{\Gamma(\alpha)}\sqrt{\frac{\alpha}{\pi\theta t}}\left(\frac{2\alpha x^2}{\theta t}\right)^{\frac{2\alpha-1}{4}}K_{\alpha-1/2}\left(\sqrt{\frac{2\alpha x^2}{\theta t}}\right),
\end{equation}
where $K_\beta(x)$ is the modified Bessel function of order $\beta$.

\subsection{Volatility autocorrelation, volatility distribution and leverage function}
The volatility  can be obtained by integrating  (\ref{eq_heston}) and is given by:
\begin{equation}
\label{eq_volprocess_int}
v_t= \left( v_0 - \theta \right) e^{-\gamma_1 t} + \theta + \kappa\int_0^{t}dW_1(u)\,e^{-\gamma(t-u)}\sqrt{v_u}. 
\end{equation} 
A simple calculation gives the stationary autocorrelation function: 
\begin{eqnarray}
\label{AC_function}
C(\tau\mid\gamma,\theta,\kappa)\equiv  \lim_{t\rightarrow\infty}\frac{\langle v_t v_{t+\tau}\rangle-\langle v_t\rangle\langle v_{t+\tau}\rangle }{\theta^2}=\frac{e^{-\gamma\tau}}{\alpha}.
\end{eqnarray}

The probability density for the volatility can be obtained as the stationary solution for a Fokker-Planck equation describing $v_t$ and reads:
\begin{equation}
\label{pdf_vol}
\Pi(v)=\frac{\alpha^\alpha}{\Gamma(\alpha)}\frac{v^{\alpha-1}}{\theta^{\alpha}}e^{-\alpha v/\theta},
\end{equation} 
implying that $\alpha>1$ is required in order to have a vanishing probability density as $v\rightarrow 0$.

The leverage function describes the correlation between returns and volatilities and is given by \cite{perello}:
\begin{equation}
\label{leverage}
L(\tau\mid\gamma,\theta,\kappa,\rho)\equiv \lim_{t\rightarrow\infty} \frac{\langle dx_t \left(dx_{t+\tau}\right)^2\rangle}{{\langle\left(dx_{t+\tau}\right)^2\rangle}^2} = \rho \kappa\, H(\tau)\, G(\tau)\,e^{-\gamma\tau},
\end{equation}
where $dx_t$ is given by (\ref{eq_detrended}), $H(\tau)$ is the Heaviside step function and: 
\begin{equation}
G(\tau)=\frac{\left\langle v_t \exp\left[\frac{\kappa}{2}\int_{t}^{t+\tau} dW_1(u)\, v_u^{-\frac{1}{2}}  \right]\right\rangle}{{\langle v_t \rangle}^2}.
\label{eq_G}
\end{equation} 
To simplify the numerical calculations  we employ throughout this paper the zeroth order appoximation $G(\tau)\approx G(0)= \theta^{-1}$.  The approximation error increases with the time lag $\tau$ but is not critical to our conclusions.

\subsection{Relation to Ehrenfest Urn Model}
\label{urn}
The Ehrenfest Urn (EU) as a model for the market return fluctuations has been studied in \cite{takahashi}, in this section we observe that Feller processes as (\ref{eq_heston}) can be produced by a sum of squared Ornstein-Uhlenbeck processes (OU), and that OU processes can be obtained as a large urn limit for the EM.
      
To see how those connections take shape we start by supposing that $X_j$ are OU processes: 
\begin{equation}
\label{eq_OU}
dX_j(t)=-\frac{b}{2}\, X_j(t)\, dt + \frac{a}{2}\, dW_j(t),
\end{equation}
where $dW_j$ describe $d$ independent Wiener processes. In this case the variable $v(t)\equiv\sum_{j=1}^{d} X_j^2(t)$ is described by a Feller process  as (\ref{eq_heston}) \cite{shreve}. The volatility process in (\ref{eq_heston}) emerges from OU processes by applying It\^o's Lemma to get:
\begin{equation}
\label{eq_volprocess_2}
dv_t=-b\;dt\sum_{j}^{d} X_j^2 +a\sum_{j}^{d}X_j\;dW_j+\frac{a^2}{4}\;\sum_{j}^{d}dW_j^2.
\end{equation}
Using the definition of $v$ and the properties of the Wiener processes it follows that:
\begin{equation}
\label{eq_volprocess}
dv_t=\left[\frac{d}{4} \;a^2-b v_t\right]\;dt \;+\;a\sqrt{v_t}\;dW.
\end{equation}
The volatility process in (\ref{eq_heston}) can be recovered with a few variable choices: $a=\kappa$,
 $b=\gamma$ and $\theta=\frac{d}{4}\frac{\kappa^2}{\gamma}$. A dynamics with vanishing probability of return to the origin requires the volatility to be represented by a sum of at least two elementary OU processes, equivalently we should have $\alpha=\frac{d}{2}\ge1$.  

An OU process can be obtained as a limit for an Ehrenfest urn model (EM). In an EM $N$ numbered balls are distributed between two urns $A$ and $B$. At each time step a number ${1,...,N}$ is chosen at random and the selected ball changes urn. We can specify $S(n)=(s_1(n),...,s_N(n))$ as the system microstate at instant $n$, where $s_j(n)=\{\pm 1\}$ indicates that the ball $j$ is inside urn $A$ if $s_j=1$,  we also can define the macrostate $M(n)=\sum_{j=1}^N s_j(n)$, which dynamics is described by a Markov chain over the space ${0,...,N}$ with transition matrix given by $Q(k,k+1)=1-k/N$, $Q(k,k-1)=k/N$ and $Q(j,k)=0$ if $|j-k|\neq1$. An imbalance in the population between the two urns generates a restitution force and, consequently, a mean reverting process for $M(n)$. Applying the thermodynamic limit $N\rightarrow \infty$, rescaling time to $t=n/N$  and introducing a rescaled imbalance variable as:
\begin{equation}
X^{(N)}_t=\sqrt{N}\left(\frac{M(tN)}{N}-\frac{1}{2}\right),
\end{equation}   
we recover an OU process:
\begin{equation}
dX_t=-2X_tdt+dW_t.
\end{equation}   

Using this connection we could speculate on a possible microscopic model that would generate a stochastic dynamics as described by the Heston model. Supposing that market agents choose at each time step between two regimes (urns) that may represent different strategies, such as technical and fundamental trading or expectations on future bullish or bearish markets, imbalances between populations in each regime would be a source of volatility. In this picture, the condition $d\ge2$ would imply that at least two independent sources of volatility would be driving the market. We shall develop this connection further elsewhere.

\section{Data Pre-processing}
\label{sec:preprocessing}

\subsection{The data}
Two data sets have been used: IB1 consisting of daily data from IBOVESPA inception on January, 1968 to January, 2005 and  IB2 consisting of high-frequency data sampled at 30 second intervals from March 3, 2001 to October 25, 2002 and from June 6, 2003 to August 18, 2004.

\begin{figure}
\begin{center}
\epsfig{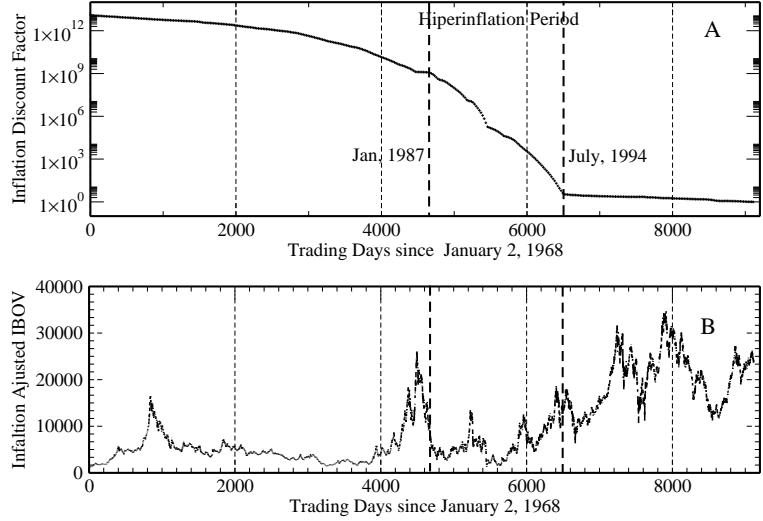}
 \caption {A: Inflation discount factor used to adjust past values to January 2005. The hyperinflation period (average of $25 \%$ per month) from January 1987 to July 1994 is also indicated. B: Inflation adjusted IBOVESPA index.} \label{data_inflation_adjust}
\end{center}
\end{figure}

\subsection{Inplits and Inflation}
\begin{figure}
\begin{center}
\epsfig{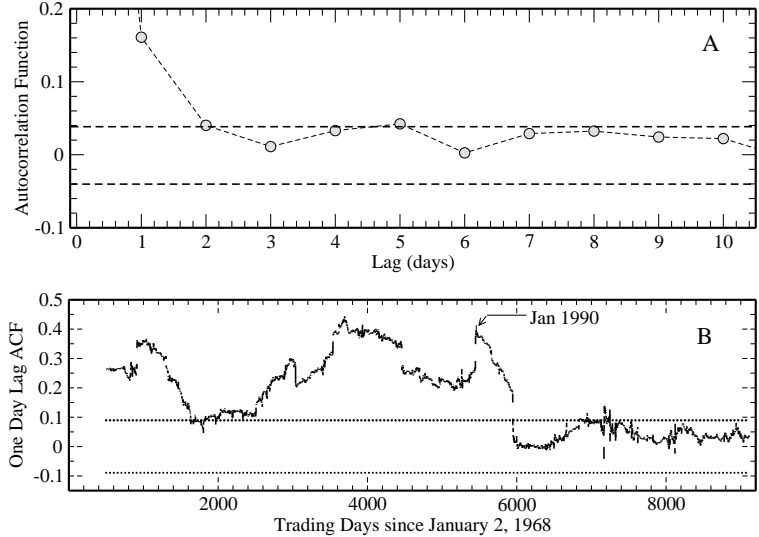}
 \caption {A: Autocorrelation function in the 1968-2005 period showing one day memory. B: Historical autocorrelation function measured with 250 days moving windows for one day lag. The behavior became compatible with a random walk after the {\it Collor Plan} in 1990.} \label{autocorr}
\end{center}
\end{figure}

The dataset IB1 has been adjusted for eleven  divisions by 10 (inplits) introduced in the period for disclosure purposes \cite{ibovespa} and also  for inflation by the General Price Index (IGP) \cite{igp}. In figure \ref{data_inflation_adjust}A we show the discount factor from date $t$ to current date $T$, $I_T(t)$, used for correcting past prices $S_t$ as  $S_t^T=S_t I_T(t)$.  Figure  \ref{data_inflation_adjust}B  shows the resulting inflation adjusted prices. The hiperinflation (average of $25\%$ per month) period from January 1987 to July 1994 is also indicated in both figures by dashed lines.

\subsection{Detrending}
For our analysis of price fluctuations it would be highly desirable to identify macroeconomic trends that may be present in the data. A popular detrending technique is the Hodrick-Prescott (HP) filtering \cite{hp} which is based on decomposing the dynamics into a permanent component, or trend $x_P(t)$, and into a stochastic stationary component $x(t)$ as $r(t)=x_P(t)+x(t)$ by fitting a smooth function to the data.
Any meaningful detrending procedure has to conserve statistical properties that define the fluctuations. However, in our experiments we have noticed that the HP filtering may introduce  subdiffusive behavior at long time scales as an artifact when applied on first differences of a random walk. 
In this paper, in the absence of a reliable detrending procedure, we  assume that the major long term trend is  represented by inflation.


\subsection{Autocorrelation of Returns}
In the period span by data set IB1 the Brazilian economy (see \cite{costa} for a brief historical account) has experienced a number of regulatory changes with consequences for price fluctuations.  In \cite{costa} it has been observed that the Hurst exponent for daily IBOVESPA returns shows an abrupt transition from an average compatible with long memory ($H>0.5$) to a random walk behavior ($H=0.5$) that coincides with major regulatory changes (known as the {\it Collor Plan}). We have confirmed  the presence of memory by measuring the autocorrelation function for the entire time series (figure \ref{autocorr}A). We have also measured the historical autocorrelation for one day lags using a 250 days moving window and confirm an abrupt behavior change coinciding with the {\it Collor Plan} in 1990 (figure \ref{autocorr}B). As the Heston model assumes  uncorrelated returns and this  feature  only became realistic after the Collor Plan,  the following analysis is restricted to a  data set consisting of daily data from January 1990 to January 2005 (IB3).

\subsection{Extreme Events and Abnormal Runs}
\begin{figure}
\begin{center}
\epsfig{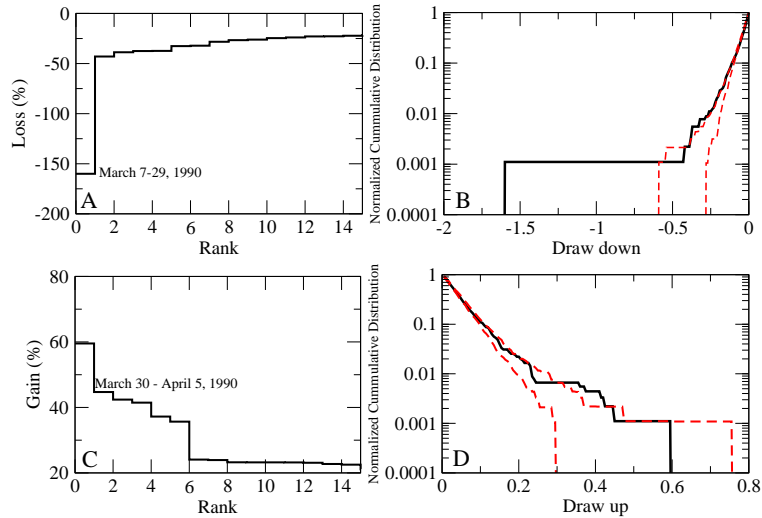}
 \caption {A and C: ranked representations of, respectively, drawdowns and drawups. B and D: full lines represent cummulative empirical distributions, dashed lines represent minimum and maximum values for 100 shuffled versions of IB3.} \label{crashes}
\end{center}
\end{figure}

Abnormal runs are sequences of returns of the same sign that are incompatible with the hypotheses of uncorrelated daily returns. We follow \cite{sornette} and calculate for the data set IB3 the statistics of persistent price decrease ({\it drawdowns}) or increase ({\it drawups}). We then compare  empirical distributions of runs with shuffled versions of IB3. In figure \ref{crashes}B we show that a seventeen business days drawdown from March 7, 1990 to March 29,1990  is statistically incompatible, within a $98\%$ confidence interval, with equally distributed  uncorrelated time series. Observe that the largest drawup shown in  figure \ref{crashes}C correponds to the subsequent period from March 30, 1990 to April 5, 1990 and  that the Collor Plan was launched in March 1990. We therefore expunged from data set IB3 abnormally correlated runs that took place in  March, 1990.

\section{Data Analysis}
\label{sec:data_analysis}

\subsection{Low Frequency}
After deflating prices and expunging autocorrelated returns, four independent parameters have to be fit  to the data: the long term mean volatility $\theta$, the relaxation time for mean reversion $\gamma^{-1}$, the amplitude of volatility fluctuations $\kappa$ and the correlation between price and volatility $\rho$. It has become apparent in \cite{silva}  that a direct least squares fit  to the probability density (\ref{px4}) yields parameters that are not uniquely defined. The data analysis adopted in this work consists in looking for statistics predicted by the model that can be easily  measured and compared. In the following subsections we describe these statistics.

\subsubsection{Long term mean volatility $\theta$}
\begin{figure}
\begin{center}
\epsfig{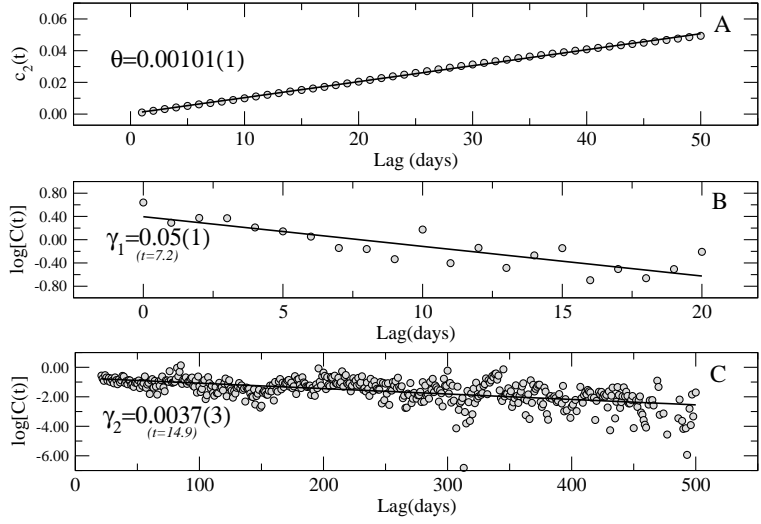}
 \caption {A: Linear regression for the second cumulant $c_2(t)$ at several time scales. The angular coefficient estimates the long-term mean volatility $\theta$. B: Linear regression for estimating the shorter relaxation time $\gamma_1^{-1}$. The $t$ statistics is shown below the estimate and assumes $18$ degrees of freedom. C: Linear regression for estimating the longer relaxation time $\gamma_2^{-1}$. The $t$ statistics assumes $445$ degrees of freedom.   } \label{corrdiff}
\end{center}
\end{figure}

A straightforward calculation yields the second cummulant of the probability density (\ref{px4}) as:
\begin{eqnarray}
\label{2cummulant}
c_2(t)&=&\left\langle x(t)^2\right\rangle-\left\langle x(t)\right\rangle^2 =-\alpha\left.\frac{\partial^2\phi_t(k)}{\partial k^2}\right|_{k=0}\\
&=&\theta t \left[1 -\epsilon\right] + \theta \frac{\epsilon}{\gamma} \left[ 1-e^{-\gamma t}\right] \nonumber, 
\end{eqnarray}
where 
\begin{equation}
\epsilon=\frac {\kappa}{\gamma}\left[\rho-\frac{1}{4}\frac{\kappa}{\gamma}\right].
\end{equation}
As $\kappa \ll \gamma$ we drop terms of $O\left( \frac{\kappa}{\gamma}\right)$ or superior. A non-biased estimate for the above cummulant can be calculated easily from data using:
\begin{equation}
	\hat{c_2}(t)=\frac{1}{N-1}\sum_{j=1}^N \left[x^{(t)}_j-\frac{1}{N}\sum_{i=1}^N x^{(t)}_i\right]^2 ,
	\label{var}
\end{equation}
where $x^{(t)}$ stands for $t$-days detrended log-returns. The long term mean volatility is estimated by a linear regression over the function $\hat{c_2}(t)$ as shown in figure \ref{corrdiff}A.  

\subsubsection{Relaxation time for mean reversion $\gamma^{-1}$}
The relaxation time is also estimated by a linear regression over the logarithm of the empirical daily volatility autocorrelation function (\ref{AC_function}) given by: 
\begin{equation}
\hat{C}(\tau) = \frac{\frac{1}{N-1}\sum_{j=1}^N  \left( x^{(1)}_{j}\right)^2 \,\left( x^{(1)}_{j+\tau}\right)^2}{\hat{\theta}^2} -1.
\end{equation}
We observe in figure \ref{corrdiff}B and  \ref{corrdiff}C  that IB3 data can be fit to two relaxation times: at $\tau<20$ we fit $\gamma_1^{-1}=20\pm4$ days and at $\tau>20$ we fit $\gamma_2^{-1}=270\pm22$ days. 

The second relaxation time can be introduced into the original Heston model by making the long-term volatility fluctuate as a mean reverting process like:
\begin{eqnarray}
\label{eq_heston_2scales}
d\theta &=& -\gamma_2\left[\theta(t)-\theta_0\right]dt\;+\;\kappa_2\sqrt{\theta(t)}\;dW_2(t),
\end{eqnarray}
where  $dW_2$ is an additional independent Wiener process. The autocorrelation function for the two time scales model acquires the following form \cite{bouchaud}:
\begin{eqnarray}
\label{AC_function_2scales}
C(\tau)= \frac{e^{-\gamma_1\tau}}{\alpha_1}+\frac{e^{-\gamma_2\tau}}{\alpha_2},
\end{eqnarray}
with $\alpha_2=\frac{2\gamma_2\bar{\theta}}{\kappa_2^2}$, where $\bar{\theta}$ stands for the average of $\theta$ given $\theta_0$. 

The autocorrelation function fit in figure \ref{corrdiff}C indicates that  relaxation times are of very different  magnitudes and that it may be possible to solve the Fokker-Planck equation for such model approximately by adiabatic elimination. We  pursue this direction further elsewhere. It is worth mentioning  that another tractable alternative for introducing multiple time scales for the volatility is a superposition of OU processes as described in \cite{barndorff}.

\subsubsection{Amplitude of volatility fluctuations $\kappa$}

From (\ref{eq_Omega}) the amplitude of volatility fluctuations is  given by:
\begin{equation}
\kappa=\sqrt{2\gamma\frac{\theta}{\alpha}}.
\label{kappa}      
\end{equation}
The long term volatility $\theta$ and the  relaxation time $\gamma^{-1}$ have been estimated in the previous sections. As discussed in section \ref{urn}, the  constant $\alpha$  is related to the number of independent OU processes composing the stochastic volatility process as $d=2\alpha$. To avoid negative volatilities, $\alpha \ge 1$ is required. We, therefore,  assume $\alpha=1$ and calculate the amplitude of volatility fluctuations from (\ref{kappa}) yielding  $\kappa=0.0032(10)$. 

\subsubsection{Correlation between prices and volatilities $\rho$}
\begin{figure}
\begin{center}
\epsfig{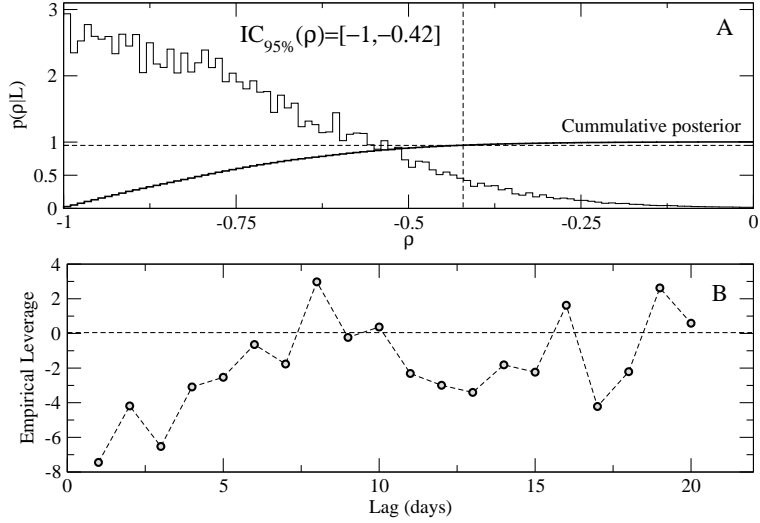}
 \caption {A: Posterior density for the correlation between prices and volatilities $\rho$ obtained by Monte Carlo sampling.  We also plot in the same graph the cummulative distribution used to compute the $95\%$ confidence interval indicated. B: Empirical leverage data as evaluated for the data set IB3.} \label{rho}
\end{center}
\end{figure}

A nonzero correlation between prices and volatilities in (\ref{eq_Wiener}) leads to an asymmetric  probability density of returns described by (\ref{px4}). This asymmetry can be estimated either by directly computing the empirical distribution skewness or by calculating the empirical leverage function (\ref{leverage}). Both procedures imply in computing highly noisy estimates for third and fourth order moments. In this section we propose  estimating a confidence interval for $\rho$ by computing the posterior probability $p(\rho\mid L)$, where $L$ corresponds to a data set containing the empirically measured leverage function for a given number of time lags. Bayes theorem \cite{sivia} gives the following posterior distribution:
\begin{equation}
p(\rho\mid L)=\frac{1}{Z(L)}\;p(\rho)\int d\gamma d\theta d\kappa \; p(L\mid \rho,\gamma,\theta,\kappa)p(\gamma)p(\theta)p(\kappa), 
\label{posterior}
\end{equation}
where $Z(L)$ is a data dependent normalization constant. We assume ignorance on the parameter $\rho$ and fix its prior distribution to be uniform over the interval $[-1,+1]$, so that $p(\rho)=U\left([-1,+1]\right)$.  The maximum entropy prior distributions  $p(\gamma)$, $p(\theta)$ and $p(\kappa)$ for the remaining parameters are gaussians, since their mean and variance have been  previously estimated. The model likelihood for time lags $0<\tau< T<\gamma_1^{-1}$ is approximately given by:
\begin{eqnarray}
\label{likelihood}
p(L\mid \rho,\gamma,\theta,\kappa)&=&\int \frac{d\sigma \,p(\sigma)}{\left(2\pi\sigma^2\right)^{\frac{T}{2}}} \\
&\times& \exp\left[-\frac{1}{2\sigma^2}\sum_{\tau=1}^T\left(\frac{1}{M\theta^2}\sum_{t=1}^M  x^{(1)}_t \,\left( x^{(1)}_{t+\tau}\right)^2-\frac{\rho \kappa}{\theta} e^{-\gamma \tau}\right)^2\right],\nonumber
\end{eqnarray}     
where the first term inside the exponencial represents the empirical leverage function with $x^{(1)}_t$ being the $M$ daily returns in the data set IB3. We choose $p(\sigma)=U\left([\sigma_{\mbox{\tiny min}},\sigma_{\mbox{\tiny max}}]\right)$ to be uniform  representing our level of ignorance on the acceptable dispersion of deviations between  data and model.  Having specified ignorance priors and the likelihood (\ref{likelihood}), we evaluate the posterior (\ref{posterior}) by Monte Carlo sampling. In figure \ref{rho} we show the resulting posterior probability density and find the $95\%$ confidence interval to be $\mbox{IC}_{\mbox{95\%}}(\rho)=[-1.00,-0.42]$, what is strong evidence for an asymmetric probability density of returns.

\subsubsection{Probability density of returns}
\begin{figure}
\begin{center}
\epsfig{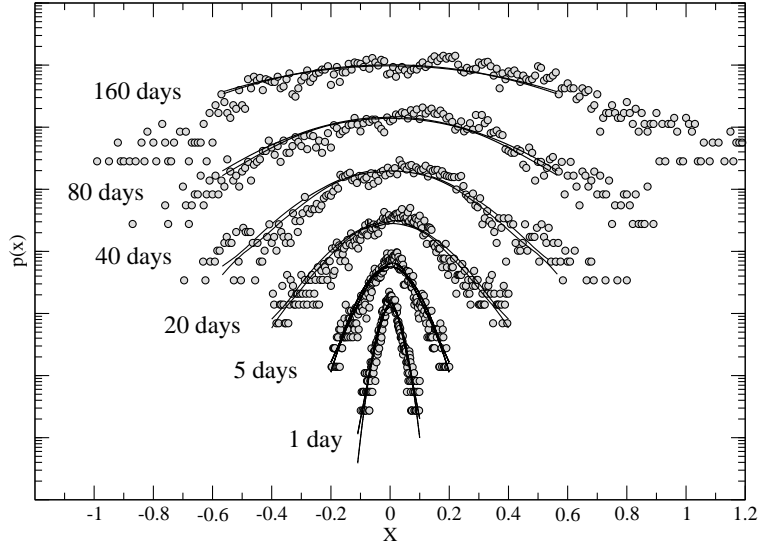}
\caption {Log-linear plot of empirical and theoretical probability densities of returns. Circles represent, from bottom to top, empirical densities at respectively $1,5,20,40,80$ and $160$ days. Full lines indicate $95\%$ confidence intervals obtained by numerical integration at Monte Carlo sampled parameter values. Densities at distinct time scales are multiplied by powers of $10$ for clarity of presentation.} \label{pdfLF}
\end{center}
\end{figure}

Confidence intervals for the  probability density of returns can be obtained by Monte Carlo sampling a sufficiently large  number of parameter sets  with appropriate distributions and numerically integrating (\ref{px4}) for each set. Distribution features for each of the relevant parameters are summarized in the following table:
\begin{center}
\begin{tabular}{|l|l|l|}\hline\hline
\emph{Parameter} & Mean & Standard Deviation \\\hline
$\theta$ & $0.00101$ $days^{-1}$ & $0.00001$ $days^{-1}$ \\\hline
$\gamma_1^{-1}$ & $20$ $days$ & $4$ $days$\\\hline
$\gamma_2^{-1}$ & $270$ $days$ & $22$ $days$\\\hline
$\kappa$ & $0.0032$ $days^{-1}$ & $0.0010$ $days^{-1}$\\\hline
$\rho$ & $\mbox{IC}_{\mbox{95\%}}=[-1.00,-0.42]$& distributed as in fig. \ref{rho}.  \\\hline
$\alpha$ & $1.0$ & fixed by theoretical arguments.\\\hline\hline
\end{tabular}
\end{center}
We have independently sampled gaussian distributions on the parameters $\theta$,$\gamma_1$,$\kappa$ and a uniform distribution $U([-1.00,-0.42])$ for the parameter $\rho$. In figure \ref{pdfLF} we compare empirical probability densities with theoretical confidence intervals at $95\%$ finding a clear agreement at time scales ranging from 1 to 160 days. 

\subsection{High Frequency}

\subsubsection{Autocorrelation of intraday returns}
\begin{figure}
\begin{center}
\epsfig{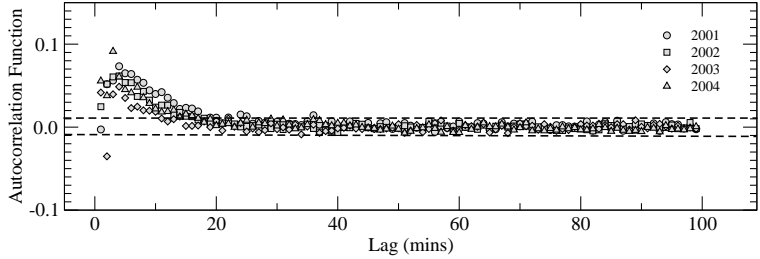}
\caption {Intraday autocorrelation function for each year composing the dataset IB2. The time scale separating micro and mesoeconomic phenomena is shown to be of about 20 minutes.} \label{autocorrHF}
\end{center}
\end{figure}

Our main aim is to describe the fluctuation dynamics at intermediate time scales of formed prices (mesosconomic time scales) by a model which assumes uncorrelated returns. The price formation  process occurs at time scales from seconds to a few minutes where the  placing of new orders and double auction processes take place. We propose to fix the shortest mesoeconomic time scale to be the point where the intraday autocorrelation function vanishes. In figure \ref{autocorrHF} we show  that the intraday return autocorrelation function vanishes at about 20 minutes for each of the 4 years composing data set IB2. We, therefore,  consider as mesoeconomic time scales over 20 minutes.    

\subsubsection{Effective duration of a day}

At first glance, it is not clear  whether intraday and daily returns can be described by the same stochastic dynamics. Even less clear is whether  aggregation from intraday to daily returns can be described by the same parameters. To verify this latter possibility we have to transform units by determining the effective duration in minutes of a business day  $T_{eff}$.  This effective duration must include the daily trading time at the S\~ao Paulo Stock Exchange  and  the impact of daily and overnight gaps over the diffusion process. The S\~ao Paulo Stock exchange opens daily for electronic trading from 10 a.m. to 5 p.m. local time and from 5:45 p.m. to 7 p.m. for after-market trading, totalizing 8h15min of trading daily.

\begin{figure}
\begin{center}
\epsfig{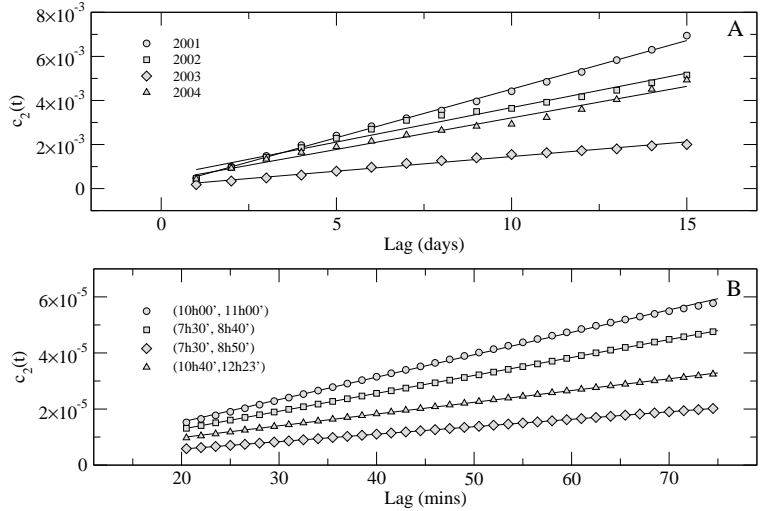}
\caption {A: Second cumulant of returns $c_2(t)$ versus the time lag in days for each year composing IB2 is used to estimate the mean daily variance $v^{(1d)}$ following the procedure described in section 4.1.1. B: Plots of intraday second cumulants $c_2(s)$ of returns versus the time lag from 20 to 80 minutes employed to estimate, via linear regression, the 20 minutes variance $v^{(20min)}$. Resulting $T_{eff}$ $95\%$ confidence intervals  are also shown in the figure. } \label{dayeff}
\end{center}
\end{figure}

To estimate  $T_{eff}$ in minutes we observe that the daily return variance $v^{(1d)}$ is the result of the aggregation of 20 minute returns, so that, considering a diffusive process, we would have:
\begin{equation}
v^{(1d)}=\frac{T_{eff}}{20} v^{(20min)}.
\label{teff}
\end{equation}
It has been already observed that the volatility fluctuation dynamics are mean reverting with at least two time scales $\gamma_1^{-1}\approx 20$ days  and  $\gamma_2^{-1}\approx 1$ year. Considering the longest relaxation time we estimate $T_{eff}$ by estimating the  mean daily volatility  for each one of the years in IB2. In figure \ref{dayeff}A we show linear regressions employed for estimating the mean daily variance $v^{(1d)}$  for each year in IB2 following the procedure described in section 4.1.1.. In figure \ref{dayeff}B we show linear regressions employed to estimate $v^{(20min)}$, the effective duration confidence intervals are obtained from (\ref{teff}). The mean $95\%$ confidence interval over the 4 years analysed results in   $IC_{95\%}\left(T_{eff}\right)=[$9h10min, 9h56min$]$ what is consistent with  8h15min of daily trading time plus an effective contribution of daily and overnight gaps. 

\subsubsection{Probability density of intraday returns}

\begin{figure}
\begin{center}
\epsfig{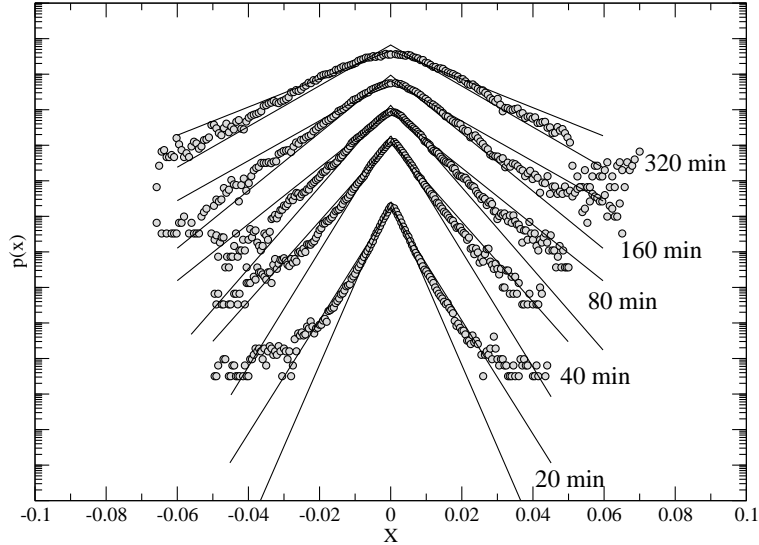}
\caption {Log-linear plot of empirical and theoretical probability densities of returns. Circles represent, from bottom to top, empirical densities at respectively $20,40,80,160$  and $320$ minutes. Full lines indicate $95\%$ confidence intervals obtained by numerical integration  at Monte Carlo sampled parameter sets rescaled by $T_{eff}$ of the short $t$ approximation for the probability density(\ref{short_t}). Densities at distinct time scales are multiplied by powers of $10$ for clarity. Note that the data shows tails fatter than predicted at shortest time scales.  } \label{pdfHF}
\end{center}
\end{figure}

For evaluating the probability density of intraday returns we have reestimated the mean volatility $\theta$  and the amplitude of volatility fluctuations $\kappa$ along  the period 2001-2004  represented in the data set IB2.   We then  have rescaled the dimensional parameters as $\theta^{(ID)}=\theta/T_{eff}$, $\gamma^{(ID)}=\gamma/T_{eff}$ and $\kappa^{(ID)}=\kappa/T_{eff}$. Having rescaled the distributions describing our ignorance on the appropriate returns we have employed Monte Carlo  sampling to compute confidence intervals for the short time lags appoximation  of the theoretical probability density described in  (\ref{short_t}). In figure \ref{pdfHF} we compare the resulting confidence intervals and the data. We attain reasonably good fits for the longer time scales, as we approach the microeconomic time scales the theoretical description of the tails breaks with the empirical data showing fatter tails.

\section{Conclusions and Perspectives}
\label{sec:conclusions}

We have studied the Heston model with stochastic volatility and exponential tails as a model for the typical price fluctuations of the Brazilian S\~ao Paulo Stock Exchange Index (IBOVESPA).  Prices have been  corrected for inflation and a period spanning the last 15 years, characterized by memoryless returns, have been chosen for the analysis. We also have expunged from data a drawdown inconsistent with the supposition of independence made by the Heston model that took place in the transition between the first 22 years of long memory returns to the memoryless time series we have analysed. 

The long term mean volatility $\theta$ has been estimated by observing the time scaling  of the log-returns variance. The  relaxation time for mean reversion $\gamma^{-1}$ has been estimated by observing the autocorrelation function of the log returns variance. We have verified that a modified version of the Heston model with two very different relaxation times ($\gamma_1^{-1}\approx 20$ days  and  $\gamma_2^{-1}\approx 1$ year) is required for describing the autocorrelation function correctly. We have used the minimum requirement for a non-vanishing volatility $\alpha \ge 1$  to calculate the scale of the variance fluctuation $\kappa$. Finally, we employed the Bayesian statistics approach for estimating a confidence interval for  the  volatility-return correlation $\rho$, as it  relies on a small data set to  calculate a  noisy estimate of higher order moments.  The quality of the model is visually inspected by comparing the empirical probability density at time scales ranging from 1 day  to 160 days, with confidence intervals obtained by a Monte Carlo simulation over the maximum ignorance parameter distributions.    We have also shown that the probability density functions of log returns at intraday time scales can be described by the Heston model with the same parameters given that we introduce an effective duration for a business day that includes the effect of overnight jumps and that we consider the slow change of the volatility due to the longer relaxation time  $\gamma_2^{-1}$.

It is surprising and non-trivial that a single  stochastic model may be capable of describing the statistical behavior of both  developed and emerging markets at a wide range of time scales, despite the known instability and high susceptibility to externalities of the latter. We believe that this robust statistical behavior may point towards simple basic mechanisms acting in the market microstructure. We regard as an interesting research direction to pursue the  derivation of the Heston model as a limit case for a microeconomic model like the minority game \cite{mg}. For this matter we believe that the connection between the Heston model and Ehrenfest urns may be valuable. 

Perhaps the search for underlying symmetries and simple basic mechanisms that can explain empirical observations should be regarded as  the main contribution of Physics  to  Economics. This contribution might be particularly useful to the field of Econometrics in which a common view  is that a theory built from data `should be evaluated in terms of the quality of the decisions that are made based on the theory' \cite{granger}. Clearly, these two approaches should not be considered as mutually exclusive.

\ack
We thank Victor Yakovenko and his collaborators for discussions and for providing useful MATLAB codes. We also wish to thank the S\~ao Paulo Stock Exchange (BOVESPA) for gently providing high-frequency data. This work has been partially (RV,VBPL) supported by {\bf FAPESP}.


\end{document}